\documentclass[12pt,usenatbib,referee]{mn2e}
\usepackage{psfig,natbib}

\def\la{\mathrel{\hbox{\rlap{\hbox{\lower4pt\hbox{$\sim$}}}\hbox{$<$}}}}
\def\ga{\mathrel{\hbox{\rlap{\hbox{\lower4pt\hbox{$\sim$}}}\hbox{$>$}}}}

\begin{document}

\bibliographystyle{mn2e}

\newcommand\aj{{AJ}}%
\newcommand\araa{{ARA\&A}}%
\newcommand\apj{{ApJ}}%
\newcommand\apjl{{ApJ}}%

\newcommand\mnras{{MNRAS}}%
\newcommand\pasp{{PASP}}%

\title
[$L^{\prime}$ and $M^{\prime}$ standard stars]
{$L^{\prime}$ and $M^{\prime}$ standard stars for the Mauna Kea
Observatories Near-Infrared (MKO--NIR) system}

\author[S.K. Leggett et al]
{S.K. Leggett,$^{1,4}$
T.G. Hawarden,$^{1,2}$
M.J. Currie,$^{1,3}$ 
A.J.Adamson,$^1$ 
T.C. Carroll,$^1$
\and
T.H. Kerr,$^1$ 
O.P. Kuhn,$^1$
M.S. Seigar,$^1$
W.P. Varricatt,$^1$
T. Wold$^1$ \\
$^1$Joint Astronomy Centre, University Park, Hilo, HI 96720, USA\\
$^2$UK Astronomy Technology Centre, Royal Observatory, Blackford Hill, 
Edinburgh EH9 3HJ, UK\\
$^3$Starlink Project, Rutherford Appleton Laboratory, Chilton, Didcot, 
Oxon OX11 0QX, UK \\
$^4$Email: {\tt skl@jach.hawaii.edu}\\
}

\maketitle
\begin{abstract}

We present $L^{\prime}$ and $M^{\prime}$ photometry,  obtained at UKIRT 
using the Mauna Kea Observatories Near--IR (MKO--NIR) filter set, for
46 and 31 standard stars, respectively.  The $L^{\prime}$ standards
include 25 from the UKIRT in--house ``Bright Standards'' with magnitudes 
deriving from Elias et al. (1982) and observations at the IRTF in the 
early 1980s, and 21 fainter stars.  The $M^{\prime}$ magnitudes derive
from the results of Sinton \& Tittemore (1984).  We estimate the average
external error to be 0$\fm$015 for the bright $L^{\prime}$
standards and 0$\fm$025 for the fainter $L^{\prime}$ standards, and
0$\fm$026 for the $M^{\prime}$ standards. The new results provide a network of
homogeneously observed standards, and establish reference stars for the
MKO system, in these bands. They also extend the available standards
to magnitudes which should be faint enough to be accessible for
observations with modern detectors on large and very large telescopes.

\end{abstract}

\section{Introduction}

Definition of an infrared photometric system was begun by Johnson and
colleagues (cf. Johnson, 1966).  At that time the $L$ and $M$ bands were
rather arbitrarily defined by quite broad filters centered at around
3.5$\mu$m and 4.8$\mu$m. Observations at $L$ employed as detectors PbS
cells, the long--wave sensitivity limits of which defined the edge of
the band, and at $M$ much less sensitive, broadband, bolometers or PbSe
cells were used.  Early infrared filters were often single remnants of 
manufacturers' batches and hence systems were not reproducible.
In addition, early infrared filter bandpasses were not well matched to the
atmospheric windows, and many absorption features, especially of water,
were included. As a consequence the exact bandpass applicable to any given
observation depended on the atmospheric water content, and observations
were not consistent between different sites or, indeed, at the same site
at different times.

The introduction of InSb detector technology made it possible to establish
a new band, centred at 3.8$\mu$m --- somewhat redwards of the $L$ band ---
which offered a much better match to the transmission of the terrestrial
atmospheric window. This has became known as the $L^{\prime}$ band, and
has effectively replaced the $L$ band, which is no longer commonly used.  
An excellent review of the status of the $LL^{\prime}M$ passbands and
existing systems, as it was in the late 1980s, is given by Bessell \& 
Brett (1988).

In the mid--1980s, as infrared detector sensitivities continued to
increase, many observatories found that the high thermal background
admitted by the broadband $M$ filter saturated the new detectors. Filters
with narrower bandpasses were introduced; these were commonly referred to
as $M^{\prime}$. Although the same labels, $L^{\prime}$ and $M^{\prime}$,
were employed at various observatories, there were often significant
differences between the particular filter bandpasses employed and hence
between their resulting photometric systems.

In the late 1990s standardization of the near--infrared filter set was
proposed by groups at the Gemini Observatories and the University of
Hawaii (Simons \& Tokunaga, 2002; Tokunaga, Simons \& Vacca, 2002). This
has been largely achieved by the design of bandpasses which are much less
sensitive to varying water vapour conditions, and the organization of a
large consortium purchase of these filters.  These filters are now known
as the Mauna Kea Observatories near--infrared (MKO--NIR) filter set.
Note, however, that they are specifically designed to allow accurate
photometry to be performed, and intercompared, at a range of altitudes,
and are not simply optimised for the Mauna Kea site.

As telescope and detector technology has continued to improve, and fainter
and cooler objects are discovered, observations in the thermal infrared have
become ever more desirable astronomically. However, standard stars for the
$L^{\prime}$ and $M$ or $M^{\prime}$ bands are still sparse and the
available set inhomogeneous. Over the last two years staff at the 3.8m UK
Infrared Telescope (UKIRT) have therefore undertaken an observing
programme to establish an enlarged and homogeneous set of photometric
standards, observed with improved accuracy through the MKO--NIR
$L^{\prime}$ and $M^{\prime}$ filters, and extending to magnitudes faint
enough to be accessible to modern (``small-well") detectors used on 
8--10-m telescopes.  This paper describes that work.  A history of the $L$
and $M$ filter bandpasses, and of the standard stars employed at UKIRT, is
given in \S 2; the observational technique in \S 3; the results in \S4;
a discussion in \S 5; and our conclusions in \S 6.

\section{UKIRT $L$ and $M$ Bandpasses and Standard Stars}

The list of standard stars in the $L$, $L^{\prime}$ and $M$ bandpasses
used by observers at UKIRT dates from 1992 and is in--house only, having
never been formally published (although at the time of writing it is 
available on the observatory
web pages). Magnitudes were derived from Elias et al. (1982) and Sinton \&
Tittemore (1984) supplemented by data from the {\em NASA Infrared Telescope
Facility Photometry Manual} (1986). UKIRT observers' own observations
between 1986 and 1989 were collected in 1989 by S. Koyonagi and T. Hawarden 
and used to correct the $J$, $H$, $K$ and $L^{\prime}$ values in the
``telescope" list. A final update based on observations in 1990--91 and
revised transformations to the CIT system was carried out by M.M. Casali in
1992. This list is known as the ``UKIRT Bright Standards".

From the 1990s on, use of the $L$ and the broadband $M$ filters ceased at
UKIRT; these were replaced by $L^{\prime}$ and a narrower--band
$M^{\prime}$ filter.  Observers simply adopted the broad--band $M$
magnitudes to calibrate their narrower--band $M^{\prime}$ data, since tests
at the telescope had shown that there was no colour dependency between the
filters, at least for spectral types earlier than K (Geballe, 1991); the
presence of photospheric CO absorption was expected to lead to differences
between $M$ and $M^{\prime}$ of a few percent for later type stars.
This is discussed further in \S 4.

UKIRT's 1--5$\mu$m imager IRCAM has a 256$\times$256 InSb detector array.  
In 1999 IRCAM was equipped with new optics, giving a smaller pixel 
field--of--view of 0$\farcs$08.  At the same time it was equipped with the 
MKO--NIR $L^{\prime}$
and $M^{\prime}$ filters (Simon \& Tokonaga, 2002;  Tokunaga et al.,
2002). Figure 1 compares the previously used ``UKIRT'' system $L^{\prime}$
bandpass with the MKO--NIR filter, and also plots transmission profiles for 
the old broadband $M$ compared with the new MKO--NIR $M^{\prime}$. 

The filter cold transmission profiles are given in Table 1, together
with the profile convolved with dry and wet atmospheric conditions as
shown in Figure 1. The atmospheric transmission spectra were obtained from 
the Gemini Observatory web pages and were calculated by Lord (1992).
The   MKO--NIR filters were designed to match the
atmospheric windows and to maximise throughput while providing better
photometric performance.  Simons \& Tokunaga (2002) calculate the
theoretical photometric error due to non--linear extinction to be
1.4 and 5.9 millimags for the  $L^{\prime}$ and $M^{\prime}$ filters,
respectively, for the Mauna Kea site.  These errors increase to 3.7
and 8.1 millimags at $L^{\prime}$ and $M^{\prime}$ for a 2~km site.  

To investigate the effect of changing water vapour content, we have 
synthesised magnitudes for the wet and dry conditions of Table 1 for G5V and 
M7V stars.  Stellar theoretical infrared spectra were obtained from Cohen 
(2003, private communication) and Hauschildt, Allard \& Baron (1999).
These calculations showed that the effect of variable water content is
small: $\leq$3 millimags at  $L^{\prime}$ and $\leq$5 millimags at  
$M^{\prime}$.

The filter profiles shown in Figure 1 and listed in Table 1 do not include
other transmission effects due to telescope and instrument optics and the 
detector.  The telescope mirrors are aluminised and light is reflected to the 
instrument  from a silver-dielectric coated dichroic tertiary mirror.  The 
camera has an uncoated calcium fluoride
window and contains coated barium fluoride 
and uncoated lithium fluoride lenses.  The detector is InSb with an 
anti--reflection coating.  Prior to installation of the MKO--NIR filters
and modification to a smaller pixel field of view, the camera had one fewer
barium fluoride and lithium fluoride lens, and two additional external gold 
mirrors.  The reflection and transmission curves of the dichroic, gold mirrors,
calcium fluoride and lithium fluoride lenses are all flat within measurement 
error over the relevant wavelength range.
The aluminised telescope  mirrors have a very small change in reflectivity 
increasing from 98.0\% at 3$\mu$m to 98.4\% at 5$\mu$m; the transmission of
the barium fluoride lens decreases from 95.6\% at 3$\mu$m to 94.4\% at 
5$\mu$m; and the array quantum efficiency (QE) varies between 87.4\% and 
90.0\%.  Calculations
of synthetic magnitudes for a G5V and an M7V star show that these optical 
elements effect
the derived magnitudes by less than 1 millimag.  To summarise, the previous
UKIRT system magnitudes, and those published here, are effectively defined 
by the filter bandpasses and the atmospheric transmission.

We note that the newer generation InSb arrays, the 1024$\times$1024
Aladdin arrays, have an anti--reflection coating with a more structured 
wavelength response.  Our current thermal imager uses an Aladdin detector
and also contains barium fluoride lenses whose coating has a 5\%--deep 
feature around 3$\mu$m.  The net change to the photometric system is 
calculated to be $\leq$1 millimag except for late--type stars at $L^{\prime}$
where the difference is $\sim$3 millimag.

We show in \S 4 that the measurement errors for the primary standards 
presented here are typically 0$\fm$015 at $L^{\prime}$ and 0$\fm$026 at 
$M^{\prime}$, therefore variations in the optical design of a 3---5$\mu$m 
imager and telescope system are unlikely to affect the photometric system to 
any measurable degree.
Furthermore, the MKO--NIR filters are designed to match the atmospheric
windows, and appear to do so well enough that, even at a lower elevation site, 
the effects of variable water vapour and non--linear extinction are also 
substantially less than the typical observational error.  Overall the results
presented here using the MKO--NIR filters should be generally useable to 
calibrate observations with conventionally designed imagers and telescopes
and on most sites.

\section{Observations}

IRCAM was used for all the observations, with the MKO--NIR $L^{\prime}$ and
$M^{\prime}$ filters.  Observations were made over 26 engineering nights
(or part--nights) between 1999 September and 2002 July. Towards the
end of this period, once most of the stars were well established, we used
calibration data from visiting observers' runs to supplement our own 
measurements. 

Primary standard stars were selected from the final UKIRT Bright Standards
list. To avoid saturation, the stars were chosen from amongst the fainter
objects on the list, with $L^{\prime} \ga 7.0$ and $M^{\prime} \ga 5.0$.  
In addition we observed new $L^{\prime}$ secondary standards  
from the UKIRT ($JHK$) Faint Standards list (Hawarden et al., 2001),
selecting stars that were expected to have $L^{\prime} \approx 10.0$. 
Typically in a full night around 30 stars were observed, 
divided equally between the three groups: bright $L^{\prime}$ primary
standards, faint $L^{\prime}$ secondary standards and bright $M^{\prime}$  
primary standards.

The IRCAM field of view is 20$\farcs$5.  Observations consisted of sets of
four or eight images with the telescope pointing slightly offset between
each; the offsets were kept small enough  that the pattern of pointings 
would fit inside the 20$\arcsec$
field.  Consecutive pairs of images were subtracted and the set of two or
four subtracted pairs were combined to give
an image containing one positive and one negative detection of the source.
Total exposure time for the bright $L^{\prime}$ standards was 80 seconds
and for the fainter stars it was typically 480 seconds, in both cases made
up of sets of four or eight 20-second exposures each in turn consisting of 100 coadded
0.2-second integrations. Exposure time at $M^{\prime}$ was usually 216
seconds, made up of sets of four or eight 9-second exposures consisting of
75 coadded 0.12-second integrations. Overheads are about a factor of two
due to the extremely short integrations. 

A flatfield was created by median filtering the set of four or eight
normalised and cleaned observations.  The cleaning removed blemishes on
scales smaller than 1 arcsec, achieved by smoothing with a box filter and
iteratively rejecting pixels more than two standard deviations from the
neighbourhood mean.  It was not necessary to mask objects when creating the
flatfield because of the extremely high sky counts in every image.  

The flatfielded subtracted images were registered using the telescope
offsets --- accurate to better than 0.5 arcsec --- converted to integer
numbers of pixels to avoid resampling.  Mosaics of the shifted
images were formed using the mean at each pixel, thus preserving flux.
Counts on and off target were dominated in all cases by the high sky 
background and
as the stars were reduced in a relative sense (described below), applying a
correction for non--linearity was not necessary.

Photometry was carried out on both the positive and negative images of the
source in the individual sky-subtracted and flatfielded frames as well as in
the final flatfielded mosaic.  The photometry aperture was automatically
positioned by the software at the source centroid; the source aperture size
used was 5$\farcs$0.  Although the background counts were close to zero,
since consecutive frames had been subtracted from one another, a concentric
``sky'' aperture was used with inner and outer radii at 6$\farcs$5 and
10$\farcs$0.  The sky value was an iteratively clipped mean, emulating the
mode.

Errors were estimated in the usual way using the variance of the
pixel-to-pixel signals in the star and background apertures.  The resulting
error estimates were consistent with the difference in the photometry
between the positive and negative image of the source, and with the scatter
of the results about the observed extinction curves (see below).  As a
further check, for a single night's observations, data variances were
created using the read noise and Poisson statistics, and propagated through
the processing steps.  The photometric errors calculated using the data
variance, yielded values some 20 percent larger for the mosaics than the
error estimates used in this paper.

Typically two to four stars were observed repeatedly (4 to 7 times) during
the night to determine the atmospheric extinction; these stars were brighter
stars chosen for their location in the sky.  These extinction stars were 
observed at airmasses as high as 2.0, typically other stars were observed
at an airmass less than 1.6.  The extinction was determined by a linear fit 
to the observed instrumental magnitude as a function of airmass.  The results
ranged from 0.08 to 0.15 magnitudes/airmass (hereinafter mag/AM) at 
$L^{\prime}$, with an average value of 0.11 mag/AM, and ranged 
from 0.18 to 0.29 mag/AM at $M^{\prime}$, with an average value 
of 0.23 mag/AM.  The error on the extinction value was on average
0.03 mag/AM at $L^{\prime}$ and 0.05 mag/AM at $L^{\prime}$.

Tokunaga et al. (2002) calculate the extinction for the MKO--NIR $L^{\prime}$ 
and  $M^{\prime}$ filters; they find that a linear fit should be accurate to 
better than 0$\fm$005.  The slope should not be strongly dependent on
the amount of water in the atmosphere: for a range
in precipitable water vapour of 0.5~mm to 4.0~mm the  $L^{\prime}$
extinction should vary from 0.09 to 0.11 mag/AM and the
$M^{\prime}$ extinction from 0.20 to 0.24 mag/AM. 
Our observed range of extinction values appears to be larger than predicted. 
However, the measurement errors are such that this this difference is not 
significant, and we note that the means of our measured values agree
with the predictions of Tokunaga et al.  No significant correlation
between our measured extinctions and the atmospheric water vapour
content is evident.

The stars not used for extinction were observed once or twice a night. Each
night's data was reduced in a relative sense: after using the extinction
stars to define a mean extinction value, the primary targets were used to
define a mean zeropoint at an airmass of unity. The fainter secondary
standards were calibrated using these values for zeropoint and extinction.
As each night's data were reduced, the deviation from the ensemble mean
zeropoint for that night of the individual zeropoints from each primary
standard was derived. This was applied as a correction to the catalogue
magnitudes for these stars, producing an evolving working catalogue to
replace the original. The evolving value was used in deriving the ensemble
zeropoint, and individual deviations, on the next night, and the process
repeated.

\section{Results}

\mbox{Table 2} gives our $L^{\prime}$ results for the stars from the UKIRT 
Bright Standards list. Spectral types for these stars were found from the 
SIMBAD database.  The number of measurements is the number of nights on which
the star was observed, i.e. if the star was observed more than once on a 
given night the mean result is counted as a single independent observation. 
Results were weighted according to the measurement error for each night.
\mbox{Table 3}
gives the $L^{\prime}$ results for the fainter stars from the UKIRT Faint
Standards of Hawarden et al. (2001), whence the spectral types are taken.
\mbox{Table 4} gives the $M^{\prime}$ results, where, again, the number of
measurements are the number of nights observed and spectral types are from
SIMBAD.  For all of Tables 2 through 4, the value of $\sigma$ given in the
sixth column is the value of the standard error of the mean over all the
nights included in the measurements; accordingly, no $\sigma$ value is
shown for stars measured on only one night. This $\sigma$ may underestimate
the true external error, as we discuss below.  

For the stars observed more than three times, any measurement that deviated 
from the mean by more than three sigma was rejected --- typically these
deviations were $\approx$0$\fm$08.  $L^{\prime}$ measurements were 
rejected for HD18881, SAO112626, HD84800, HD161903, FS104, FS123, FS125, 
FS147, FS149 and FS155.  $M^{\prime}$ measurements were rejected for 
HD84800 and HD129653.  In all cases single measurements were discarded,
except for SAO112626 and FS155 where two  $L^{\prime}$ measurements 
were discarded.  
We interpret these deviant values as simply bad data and not as
evidence of variability at $L^{\prime}$ or $M^{\prime}$. All the stars
have been measured on several
nights at near--infrared and, in most cases, optical wavelengths. The bright
(primary) stars were observed in the $J$, $H$, $K$ and $L$ bands on
numerous occasions at UKIRT (see above) and at CTIO and KPNO (Elias et
al., 1982) over more than a decade. The faint stars have been observed at
$JHK$ an average of 7 (minimum: 3, maximum: 12) times over the period
1994--1998 (Hawarden et al. 2001) and in two cases are Landolt (1992)
standards with numerous observations in the optical. None have been
suspected to vary. Since stellar variability is usually larger in
amplitude at shorter wavelengths (barring large changes in atmospheric
structure, such as dust formation, unlikely in stars as carefully selected
for stability and normality as these) changes of significant amplitude
must be regarded as unlikely.  An exception to this may be FS 101, for
which the error in the mean $L^{\prime}$ value remained high despite repeat 
measurements.  This F0 star did not show any signs of variability at
$JHK$ (Hawarden et al. 2001) but seven  $L^{\prime}$ measurements showed a 
range of 0$\fm$1 with a standard error in the mean of 0$\fm$05.  We note
that the sample includes two B supergiants, BS 696 and BS 8541, as 
$M^{\prime}$ standards.  Although it is possible that free--free emission 
from potentially variable stellar winds contributes to their  $M^{\prime}$ 
flux there is currently no evidence of variability. Sinton \& Tittemore 
(1984) observed these stars 6--7 times and they have also been observed 
numerous times at UKIRT (BS 8541 is flagged as a frequently observed star 
in the UKIRT Bright Standards list). Note that if there is a significant
contribution from a stellar wind then the stars will have atypical colours
and should not be used for system transformations.

\mbox{Table 5} combines all the measurements in one table, and also lists 
Right Ascension and Declination, proper motion and spectral type.  Here we 
have estimated the external errors as follows.  On individual nights
the scatter around the mean extinction curve for stars observed more
than once was typically 0$\fm$02 for the bright $L^{\prime}$ standards, and 
0$\fm$03 for the fainter $L^{\prime}$ standards and for the $M^{\prime}$ 
standards. This is only slightly worse than the average calculated error 
of an individual measurement, based on signal--to--noise statistics. 
Accordingly, if the 
star was observed on $N$ nights then our adopted uncertainty in 
\mbox{Table 5} is 
the larger of $0.02[3]/ \sqrt(N-1)$ or the internal standard deviation of 
the mean given in Tables 1--3.  

The number of observations in \mbox{Table 5} is the number of nights on
which the star was observed in each filter. A combination of lack of
photometric weather and the replacement of IRCAM by another instrument
necessitated the termination of this programme when some stars had fewer
than three observations.  These stars, and those with 
estimated error larger than 0$\fm$03, should be treated with 
appropriate caution, as perhaps should the two B supergiants mentioned above.

\section{Discussion}

Figure 2 plots, for the brighter stars, the difference between our
$L^{\prime}$ and $M^{\prime}$ magnitudes on the MKO system and the
previously listed values on the UKIRT $L^{\prime}$ and broadband $M$
systems. The differences are plotted as a function of colour, expressed as
$J - K$ on the UKIRT system (not MKO), and of brightness at $L^{\prime}$
or $M^{\prime}$.

At $L^{\prime}$ the mean absolute difference between the current and
previous measurements is 0$\fm$02, consistent with our uncertainties and
the estimated error in the older measurements (e.g. the two stars with 
$L^{\prime} \approx 6$ in Sinton \& Tittemore (1984) have a quoted 
uncertainty of 0$\fm$02 and 0$\fm$05 in their \mbox{Table 2}). There is
no evidence of a colour term; instead the stars with the largest
deviations from the previously tabulated value are fainter, suggesting
larger errors in the original measurements.  The latest spectral type in 
this sample is M5.5.  It is known that for ultracool objects, of spectral type
L5 and later, there will be a difference in magnitude measured with the
previous UKIRT $L^{\prime}$ and the current MKO--NIR $L^{\prime}$ filter.  
This is due to the onset of methane absorption at the blue edge of the 
MKO--NIR bandpass (Leggett et al., 2002). Such objects are not currently used
as standards.

At $M^{\prime}$ the mean absolute difference between the current
and previous measurements is 0$\fm$05, again consistent with our
estimated errors and those of the older measurements (e.g. the two stars with
$M^{\prime} \approx 6$ in Sinton \& Tittemore (1984) have quoted errors of
0$\fm$03 and 0$\fm$06). Again, the deviation is larger for fainter
targets suggesting a larger error in the original measurement.  There are 
insufficient red stars in common to determine whether or not a colour term 
exists between $M$ and $M^{\prime}$ magnitudes.  Although a dependency was 
expected for types K and later due to photospheric CO absorption, we have 
synthesised broadband 
$M$ and $M^{\prime}$ magnitudes for a late--M dwarf and for an early--K giant
(using models from  Hauschildt, Allard \& Baron (1999) for the former and the
observationally based templates of Cohen et al. (1999) for the latter) and 
found the difference to be only 0$\fm$005, presumably because both filters
include the CO band. In any case, as broadband $M$ is no longer used this is
not an issue for modern work.

\section{Conclusions}

We have presented new $L^{\prime}$ observations, using the MKO--NIR filter, of
25 stars with $L^{\prime} \sim 7$.  The average internal standard error of
the mean result for each star is 0$\fm$010, the estimated external 
error ranges from 0$\fm$010 to 0$\fm$025.  Most of these stars have 
previously tabulated $L^{\prime}$ values on the ``UKIRT'' system.  
The difference between the UKIRT and MKO values are typically 0$\fm$02, 
with one larger deviation of 0$\fm$09 for the faintest star in the sample.  
There is no evidence of a colour dependency in 
$L^{\prime}_{MKO} - L^{\prime}_{UKIRT}$ for stars as late as M5.5.

Data with the same filter are presented for 21 stars with $L^{\prime} \sim
10$, taken from the UKIRT $JHK$ Faint Standards list. The average internal
standard error of the mean results is 0$\fm$02, the estimated external error
ranges from 0$\fm$02 to 0$\fm$05. These stars will be useful as standards for 
larger telescopes, and convenient for programs on such telescopes which 
require calibration of all four of the $JHKL^{\prime}$ bands.

We also present $M^{\prime}$ observations, using the MKO--NIR filter, of 31
stars with $M^{\prime} \sim 6.5$.  The average internal standard error of
the mean results is 0$\fm$02, while the estimated external error ranges from 
0$\fm$02 to 0$\fm$05.  Most of these stars have previously tabulated 
broadband $M$ magnitudes; and the difference between $M$ and MKO--NIR
$M^{\prime}$ is typically 0$\fm$05 although there are two stars that
differ by $>$0$\fm$1; these are fainter stars that most likely have larger
uncertainties in the original measurements.  The new data will allow more
accurate calibration of $M$--band photometry on a wide range of
telescopes.

Systematic errors in the photometry due to non--linear extinction or 
variable water vapour are calculated to be  $\leq$4 millimags at  $L^{\prime}$ 
and $\leq$8 millimags at $M^{\prime}$, for sites as low as 2~km.
Investigation of the transmissive or reflective elements of UKIRT and its
imagers implies that commonly used optical elements will introduce
variations in the photometric system of $\leq$3 millimag.
Hence the results presented here are generally applicable to other 
observatories.

\section*{Acknowledgments}

UKIRT, the United Kingdom Infrared Telescope, is operated by the Joint
Astronomy Centre on behalf of the U.K. Particle Physics and Astronomy
Research Council (PPARC). This work would not have been possible without the 
dedicated effort of all the UKIRT staff, past and present; in particular 
we note the important
contribution of the late Sidney Arakaki, without which none of the early
UKIRT photometry would have been possible. We are very grateful to
Mark Casali, Tim Chuter, Maren Hauschildt--Purves, Kevin Krisciunas, Andy
Longmore, Erik Starman, Mike Wagner and Peredur Williams, as well as
vacation student Stuart K. Koyonagi. We thank visiting observers Dave
Golimowski and Gene Magnier for allowing us to incorporate their
calibration observations in these results.  We also thank the referee
for a careful reading of the manuscript and for suggestions that lead to 
substantial improvements.

ORAC--DR was used to reduce the observations for this paper.  ORAC--DR was
developed at the  Joint Astronomy Centre; the concept and early recipes
originated at the  Astronomy Technology Centre, Edinburgh.  The 
application engines used in ORAC--DR were 
supplied and excellently supported by the Starlink Project, which is run by 
the UK Central Laboratory for the Research Councils on behalf of PPARC.  We 
thank all the programmers involved.


\newpage

\begin{table}
\caption{Profiles of the MKO--NIR
$L^{\prime}$ and $M^{\prime}$ filters at cold temperatures, 
and including the effect
of absorption by an atmosphere with 1.2~mm and 3~mm of water vapour. }
\begin{tabular}{rrrrrrrr}
\multicolumn{4}{c}{$L^{\prime}$} & \multicolumn{4}{c}{$M^{\prime}$} \\
Wavelength  & Transmission & \multicolumn{2}{c}{H$_2$O}  &
Wavelength  & Transmission &  \multicolumn{2}{c}{H$_2$O} \\
$\mu$m &  \% & 1.2mm  & 3.0mm  &
$\mu$m & \% & 1.2mm  & 3.0mm  \\

\hline
3.22   &  0.0  &  0.0  &  0.0 &  4.38  &  0.0  &  0.0  &  0.0\\
3.24   &  0.1  &  0.1  &  0.1  & 4.40  &  0.2  &  0.0 &   0.0\\
3.26    & 0.1  &  0.1  &  0.1 &  4.42  &  0.2  &  0.0  &  0.0\\
3.28   &  0.2  &  0.2  &  0.1 &  4.44  &  0.5  &  0.0  &  0.0\\
3.30   &  0.6  &  0.4   & 0.3  & 4.46  &  0.3  &  0.0  &  0.0\\
3.32   &  1.1   & 0.5  &  0.4  & 4.48  &  0.3  &  0.1  &  0.0\\
3.34   &  1.9   & 1.5  &  1.3  & 4.50  &  0.7  &  0.3  &  0.3\\
3.36   &  3.1   & 2.4  &  2.1  & 4.52   & 2.0  &  0.6  &  0.6\\
3.38   &  6.1  &  4.6  &  4.2 &  4.54  &  8.6  &  3.6  &  3.3\\
3.40   & 16.2  & 12.2  & 11.6  & 4.56  & 36.8  & 21.3  & 20.0\\
3.42   & 43.2  & 37.1  & 36.5 &  4.58  & 84.4  & 58.6  & 54.6\\
3.44   & 76.7  & 74.1  & 72.1  & 4.60  & 91.3 &  73.8  & 70.2\\
3.46  &  90.1  & 85.9  & 83.1 &  4.62  & 92.0  & 78.3 &  75.5\\
3.48  &  91.8  & 87.9  & 86.1  & 4.64  & 93.2  & 80.0 &  77.0\\
3.50  &  91.6  & 89.5  & 87.6 &  4.66  & 91.5 &  80.2  & 73.2\\
3.52   & 89.4  & 87.8  & 85.8  & 4.68 &  89.4  & 73.8 &  65.2\\
3.54   & 87.6 &  80.9  & 78.6 &  4.70  & 88.8 &  66.8  & 66.2\\
3.56   & 87.9  & 84.0  & 81.7 &  4.72  & 89.6  & 64.7 &  62.9\\
3.58   & 89.6 &  84.4 &  81.4 &  4.74  & 90.8  & 72.1 &  70.2\\
3.60  &  91.4  & 87.8 &  85.4  & 4.76  & 89.1 &  69.3  & 67.2\\
3.62  &  92.7 &  89.6  & 87.5  & 4.78 &  78.0 &  59.4  & 52.4\\
3.64  &  93.6  & 90.6  & 89.0  & 4.80  & 53.9  & 44.1  & 40.0\\
3.66  &  94.1  & 91.5 &  90.0 &  4.82  & 25.6 &  19.6  & 17.4\\
3.68  &  93.6  & 89.3  & 84.5  & 4.84  &  9.1   & 5.4  &  4.4\\
3.70   & 92.8 &  90.3  & 89.1 &  4.86  &  3.0  &  2.4  &  2.3\\
3.72  &  91.5  & 88.3  & 86.8  & 4.88  &  1.1   & 1.0  &  1.0\\
3.74  &  90.6 &  86.9  & 85.8  & 4.90  &  0.5  &  0.4  &  0.2\\
3.76  &  91.0  & 86.7 &  84.4  & 4.92  &  0.5   & 0.5  &  0.5\\
3.78  &  91.7  & 88.1  & 86.2  & 4.94   & 0.0  &  0.0  &  0.0\\
3.80  &  92.1  & 87.7  & 86.2 & & & & \\
3.82  &  92.7  & 88.4 &  86.6 & & & & \\
3.84  &  92.8  & 87.4  & 85.5  & & & & \\
3.86  &  92.7  & 86.0  & 84.2 & & & & \\
3.88  &  93.1  & 80.4  & 78.1  & & & & \\
3.90  &  93.6  & 84.2  & 82.5 & & & & \\
3.92  &  94.1  & 81.9 &  79.8 & & & & \\
3.94  &  93.7  & 83.3 &  81.8 & & & & \\
3.96  &  92.5  & 83.3  & 82.0 & & & & \\
3.98  &  90.8 &  80.4  & 79.2 & & & & \\
4.00  &  89.7  & 77.7  & 76.3 & & & & \\
4.02  &  90.2  & 75.6  & 74.0 & & & & \\
4.04  &  91.7  & 73.2  & 71.4 & & & & \\
4.06  &  93.7  & 72.6  & 70.6 & & & & \\
4.08 &   92.2  & 67.5  & 65.2 & & & & \\
4.10  &  79.3  & 54.4 &  52.3 & & & & \\
4.12  &  53.3 &  30.5 &  29.0 & & & & \\
4.14 &   27.2  & 10.5  &  9.9 & & & & \\
4.16 &   10.6  &  2.5  &  2.3 & & & & \\
4.18  &   4.2  &  0.3  &  0.3 & & & & \\
4.20  &   1.9  &  0.1  &  0.0 & & & & \\
4.22   &  0.8  &  0.0  &  0.0 & & & & \\
4.24  &   0.4  &  0.0  &  0.0 & & & & \\
4.26  &   0.0  &  0.0  &  0.0 & & & & \\
\end{tabular}
\end{table}

\begin{table}
\caption{New $L^{\prime}$ Photometry for Bright Standards}
\begin{tabular}{llclrrc}
Name & Other & RA/Dec & Spectral & $L^{\prime}$  & $\sigma$ & Number of \\
  & Names  & equinox 2000 & Type & mag  & mag & Observations\\
\hline
HD225023  &  SAO53596    & 00:02:46.03 $+$35:48:55.7 &   A0   &  6.979 
& 0.001 &  3\\
G158-27   &  GJ1002      & 00:06:43.00 $-$07:32:42.0 &   M5.5V &  6.989 
& 0.018 &  4\\
HD1160    &  SAO109094   & 00:15:57.30 $+$04:15:04.0 &   A0   &  7.055 
& 0.005 &  3 \\ 
HD3029    &  SAO74098    & 00:33:39.53 $+$20:26:01.7 &   A3   &  7.082 
& 0.014 &  4\\
HD18881   &  SAO56114    & 03:03:31.94 $+$38:24:36.1 &   A0   &  7.160 
& 0.007 &  3\\
HD22686   &  SAO111318   & 03:38:55.09 $+$02:45:48.6 &   A0   &  7.199 
& 0.008 &  4\\
SAO112626 &  HD287736    & 05:19:17.16 $+$01:42:16.1 &   G0   &  8.559 
& 0.010 &  3\\
HD38921   &  SAO196174   & 05:47:22.19 $-$38:13:51.3 &   A0V   &  7.513 
& 0.013 &  2\\
HD40335   &  SAO113311   & 05:58:13.52 $+$01:51:23.0 &   A0   &  6.441 
& 0.025 &  2 \\
HD44612   &  SAO41080    & 06:24:46.60 $+$43:32:54.5 &   A0   &  7.050 
& 0.002 &  3\\
HD77281   &  SAO136505   & 09:01:38.01 $-$01:28:34.8 &   A2   &  7.041 
& 0.014 &  6\\
GL347A    &  G161-33     & 09:28:53.50 $-$07:22:15.0 &   M2.5V &  7.367 
& 0.009 &  3 \\
HD84800   &  SAO43050    & 09:48:44.64 $+$43:39:55.6 &   A2II   &  
7.547 & 0.013 &  5 \\
HD105601  &  SAO62866    & 12:09:27.80 $+$38:37:54.6 &   Am    &  6.669 
& 0.011 &  3 \\ 
HD106965  &  SAO119313   & 12:17:57.54 $+$01:34:31.1 &   A2   &  7.311 
& 0.010 &  6\\
HD129653  &  SAO64289    & 14:42:39.56 $+$36:45:24.3 &   A2   &  6.920 
& 0.007 &  3 \\ 
HD129655  &  SAO140097   & 14:43:46.44 $-$02:30:20.0 &   A2   &  6.666 
& 0.014 &  3\\  
HD136754  &  SAO83785    & 15:21:34.53 $+$24:20:36.1 &   A0   &  7.158 
& 0.010 &  5\\
HD162208  &  SAO66344    & 17:47:58.56 $+$39:58:50.9 &   A0   &  7.125 
& 0.011 &  3 \\ 
HD161903  &  SAO141886   & 17:48:19.22 $-$01:48:29.7 &   A2   &  7.034 
& 0.005 &  3 \\ 
HD161743  &  SAO209292   & 17:48:57.93 $-$38:07:07.5 &   B9IV   &  
7.623 & 0.001  &  2\\
GL748     &  G22-18      & 19:12:14.60 $+$02:53:11.1 &   M3.5V &  6.012 
& 0.017  &  2\\
GL811.1   &  Wolf 896    & 20:56:46.60 $-$10:26:54.6 &   M2.5V &  6.691 
& 0.006 &  3\\ 
HD203856  &  SAO71278    & 21:23:35.53 $+$40:01:07.0 &   A0   &  6.871 
& 0.013 &  5\\
SAO34401  &  HD212533    & 22:23:42.24 $+$55:12:25.1 &   F0V   &  7.735 
& 0.013 &  6\\
\end{tabular}
\end{table}

\begin{table}
\caption{New $L^{\prime}$ Photometry for Faint Standards}
\begin{tabular}{llclrrc}
UKIRT FS & Other & RA/Dec & Spectral & $L^{\prime}$  & $\sigma$ & Number of\\
Number  & Names  & equinox 2000 & Type & mag  & mag & Observations\\
\hline
101 & CMC400101   &    00:13:43.58 $+$30:37:59.9  & F0   & 10.34  & 0.05 & 7 \\
2   & SA92-342 &    00:55:09.93 $+$00:43:13.1  & F5   & 10.44  & 0.02 & 4 \\                
104 & P194-R   &    01:04:59.43 $+$41:06:30.8  & A7   & 10.36  & 0.03 & 3 \\                
107 & CMC600954   &    01:54:10.14 $+$45:50:38.0  & G0   & 10.18  & 0.02 & 3 \\ 
108 & CMC502032   &    03:01:09.85 $+$46:58:47.7  & F8   &  9.65  & 0.01 & 3 \\
109 & LHS169    &    03:13:24.16 $+$18:49:38.4  & M2V   & 10.50  & 0.01 & 3 \\
111 & CMC601790   &    03:41:08.55 $+$33:09:35.5  & G5   & 10.23  & 0.01 & 3 \\
117 & B216-b9  &    04:23:56.61 $+$26:36:38.0  & N/A  &  9.75  & 0.01 & 2 \\                
119 & SAO131719   &    05:02:57.44 $-$01:46:42.6  & A2   &  9.80  & 0.01 & 2 \\   
13  & SA97-249 &    05:57:07.59 $+$00:01:11.4  & G4   & 10.10  & 0.03 & 2 \\  
123 & P486-R   &    08:51:11.88 $+$11:45:21.5  & B8   & 10.25  & 0.02 & 4 \\ 
125 & P259-C   &    09:03:20.60 $+$34:21:03.9  & G8   & 10.33  & 0.03 & 3 \\ 
129 & LHS2397a &    11:21:48.95 $-$13:13:07.9  & M8V   & 10.03  & N/A  & 1 \\
134 & LHS2924  &    14:28:43.37 $+$33:10:41.5  & M9V   & 10.10  & 0.02 & 3 \\ 
138 & P275-A   &    16:28:06.72 $+$34:58:48.3  & A1   & 10.44  & 0.03 & 3 \\                 
140 & S587-T   &    17:13:22.65 $-$18:53:33.8  & G9   & 10.34  & 0.01  & 2 \\
147 & P230-A   &    19:01:55.27 $+$42:29:19.6  & A0   &  9.84  & 0.02 & 3 \\
148 & S810-A   &    19:41:23.52 $-$03:50:56.1  & A0   &  9.46  & 0.02 & 4 \\                
149 & P338-C   &    20:00:39.25 $+$29:58:40.0  & B7.5 & 10.06  & 0.02 & 5 \\
150 & CMC513807   &    20:36:08.44 $+$49:38:23.5  & G0   &  9.91  & 0.02 & 4 \\
155 & CMC516589    &    23:49:47.82 $+$34:13:05.1  & K5   &  9.32  & 0.02 & 5 \\
\end{tabular}
\end{table}

\begin{table}
\caption{New $M^{\prime}$ Photometry for Standards}
\begin{tabular}{llclrrc}
Name & Other & RA/Dec & Spectral & $M^{\prime}$  & $\sigma$ & Number of\\
  & Names  & equinox 2000 & Type & mag  & mag & Observations\\
\hline
HD225023  &  SAO53596    & 00:02:46.03 $+$35:48:55.7 &   A0   &  6.95 & 0.03 &  8 \\
G158-27   &  GJ1002      & 00:06:43.00 $-$07:32:42.0 &   M5.5V &  7.03 & 0.05 &  2 \\
HD1160    &  SAO109094   & 00:15:57.30 $+$04:15:04.0 &   A0   &  7.04 & 0.01 &  3 \\ 
HD3029    &  SAO74098    & 00:33:39.53 $+$20:26:01.7 &   A3   &  7.04 & 0.02 &  3 \\
BS696     &  HD14818     & 02:25:16.03 $+$56:36:35.4 &   B2Iae   &  5.32 & 0.03 &  3 \\
HD18881   &  SAO56114    & 03:03:31.94 $+$38:24:36.1 &   A0   &  7.17 & 0.02 &  3 \\
HD22686   &  SAO111318   & 03:38:55.09 $+$02:45:48.6 &   A0   &  7.16 & 0.02 &  4 \\
BS1140    &  HD23288     & 03:44:48.22 $+$24:17:22.1 &   B7IV  &  5.57 & 0.02 &  3 \\
BS1869    &  HD36719     & 05:36:15.96 $+$47:42:55.0 &   F0V   &  5.40 & N/A  &  1 \\
HD38921   &  SAO196174   & 05:47:22.19 $-$38:13:51.3 &   A0V   &  7.49 & 0.02 &  2 \\
HD40335   &  SAO113311   & 05:58:13.52 $+$01:51:23.0 &   A0   &  6.41 & 0.01 &  5 \\
BS2228    &  HD43244     & 06:17:34.65 $+$46:25:26.2 &   F0V   &  5.87 & 0.02 &  2 \\ 
HD44612   &  SAO41080    & 06:24:46.60 $+$43:32:54.5 &   A0   &  7.07 & 0.03 &  3 \\
HD77281   &  SAO136505   & 09:01:38.01 $-$01:28:34.8 &   A2   &  7.02 & 0.03 &  5 \\
HD84800   &  SAO43050    & 09:48:44.64 $+$43:39:55.6 &   A2II   &  7.56 & 0.01 &  5 \\
HD105601  &  SAO62866    & 12:09:27.80 $+$38:37:54.6 &   Am    &  6.70 & 0.02 &  3 \\ 
HD106965  &  SAO119313   & 12:17:57.54 $+$01:34:31.1 &   A2   &  7.32 & 0.04 &  3 \\
HD129653  &  SAO64289    & 14:42:39.56 $+$36:45:24.3 &   A2   &  6.99 & 0.01 &  3 \\ 
HD129655  &  SAO140097   & 14:43:46.44 $-$02:30:20.0 &   A2   &  6.69 & 0.02 &  3 \\  
HD136754  &  SAO83785    & 15:21:34.53 $+$24:20:36.1 &   A0   &  7.13 & 0.04 &  4 \\
BS6092    &  HD147394    & 16:19:44.44 $+$46:18:48.1 &   B5IV   &  4.37 & 0.01  &  2 \\
HD162208  &  SAO66344    & 17:47:58.56 $+$39:58:50.9 &   A0   &  7.05 & 0.02 &  3 \\ 
HD161903  &  SAO141886   & 17:48:19.22 $-$01:48:29.7 &   A2   &  6.97 & 0.01 &  3 \\ 
HD161743  &  SAO209292   & 17:48:57.93 $-$38:07:07.5 &   B9IV   &  7.67 & N/A  &  1 \\
GL748     &  G22-18      & 19:12:14.60 $+$02:53:11.1 &   M3.5V &  6.00 & 0.01  &  2 \\
BS7773    &  HD193432    & 20:20:39.82 $-$12:45:32.7 &   B9IV   &  4.86 & 0.02 &  3 \\
GL811.1   &  Wolf 896    & 20:56:46.60 $-$10:26:54.6 &   M2.5V &  6.72 & 0.02 &  3 \\ 
HD201941  &  SAO126618   & 21:12:45.32 $+$02:38:33.9 &   A2   &  6.63 & 0.03 &  3 \\
HD203856  &  SAO71278    & 21:23:35.53 $+$40:01:07.0 &   A0   &  6.84 & 0.01 &  3 \\
SAO34401  &  HD212533    & 22:23:42.24 $+$55:12:25.1 &   F0V   &  7.70 & 0.02 &  3 \\
BS8541    &  HD212593    & 22:24:30.99 $+$49:28:35.0 &   B9Iab   &  4.20 & 0.01 &  3 \\
\end{tabular}
\end{table}

\begin{table}
\caption{Summary of $L^{\prime}$$M^{\prime}$ Photometry}
\begin{tabular}{llcrrlrlrcc}
Name & Other & RA/Dec & \multicolumn{2}{c}{Proper Motion} & Type & $L^{\prime}$ & Estimated &  
$M^{\prime}$ & Estimated & Number\\
 & Name  & equinox 2000 & \multicolumn{2}{c}{mas/year} & &  mag  & Error & mag & Error  & of Obs.\\
\hline
HD225023  &  SAO53596    & 00:02:46.03 $+$35:48:55.7 & $+$14 & $-$2 &  A0   &  6.979 
& 0.014 &  6.95 & 0.03 & 3,8\\
G158-27   &  GJ1002      & 00:06:43.00 $-$07:32:42.0 & $-$623 & $-$2037 &  M5.5V &  6.989 
& 0.018 &  7.03 & 0.05 & 4,2 \\
FS101     &  CMC400101      & 00:13:43.58 $+$30:37:59.9 & $-$5 & $-$9 &  F0   & 
10.34\phantom {0} & 0.05 &       &   & 7,0   \\
HD1160    &  SAO109094   & 00:15:57.30 $+$04:15:04.0 &  $+$21 & $-$14 &  A0   &  7.055 
& 0.014 &  7.04 & 0.02 &   3,3\\ 
HD3029    &  SAO74098    & 00:33:39.53 $+$20:26:01.7 & $+$4 & $+$1 &   A3   &  7.082 
& 0.014 &  7.04 & 0.02 & 4,3\\
FS2       &  SA92-342    & 00:55:09.93 $+$00:43:13.1 & &  & F5   & 
10.44\phantom {0} & 0.02 &       &   & 4,0   \\ 
FS104     &  P194-R      & 01:04:59.43 $+$41:06:30.8 & $+$0 & $-$4 &  A7   & 
10.36\phantom {0} & 0.03 &       &   & 3,0   \\ 
FS107     &  CMC600954    & 01:54:10.14 $+$45:50:38.0 &  $-$25 & $-$4 & G0   & 
10.18\phantom {0} & 0.02 &       &  & 3,0    \\ 
BS696     &  HD14818     & 02:25:16.03 $+$56:36:35.4 &  $-$0  & $-$1 &  B2Iae   &       &    
&  5.32 & 0.03 & 0,3\\
FS108     &  CMC502032   & 03:01:09.85 $+$46:58:47.7 &  $+$1  & $-$1 &  F8   &  
9.65\phantom {0} & 0.02 &       &   & 3,0   \\
HD18881   &  SAO56114    & 03:03:31.94 $+$38:24:36.1 & $+$5  & $-$12 &   A0   &  7.160 
& 0.014 &  7.17 & 0.02 & 3,3\\
FS109     &  LHS169      & 03:13:24.16 $+$18:49:38.4 &  $+$1346 &  $-$1103 & M2V   & 
10.50\phantom {0} & 0.02 &       &    & 3,0  \\
HD22686   &  SAO111318   & 03:38:55.09 $+$02:45:48.6 & $+$24 &  $-$20 &   A0   &  7.199 
& 0.012 &  7.16 & 0.02 & 4,4\\
FS111     &  CMC601790      & 03:41:08.55 $+$33:09:35.5 &  $+$3 &  $+$3 &  G5   & 
10.23\phantom {0} & 0.02 &       &   & 3,0   \\
BS1140    &  HD23288     & 03:44:48.22 $+$24:17:22.1 &   $+$21 &  $-$44 &  B7IV   &      &      
&  5.57 & 0.02 & 0,3\\
FS117     &  B216-b9     & 04:23:56.61 $+$26:36:38.0 &  &  & N/A  &  
9.75\phantom {0} & 0.03 &       &   & 2,0   \\  
FS119     &  SAO131719       & 05:02:57.44 $-$01:46:42.6 & $+$1 & $-$6 &   A2   &  
9.80\phantom {0} & 0.03 &       &  & 2,0  \\
SAO112626 &  HD287736    & 05:19:17.16 $+$01:42:16.1 & $+$22&  $-$41 &   G0   &  8.559 
& 0.010 &       &   & 3,0   \\
BS1869    &  HD36719     & 05:36:15.96 $+$47:42:55.0 & $+$14&  $-$20 &   F0V   &       &      
&  5.40 & 0.03 & 0,1\\
HD38921   &  SAO196174   & 05:47:22.19 $-$38:13:51.3 &  $+$0 & $-$7 &  A0V   &  7.513 
& 0.020 &  7.49 & 0.03 & 2,2\\
FS13      &  SA97-249    & 05:57:07.59 $+$00:01:11.4 & & &   G4   & 
10.10\phantom {0} & 0.03 &       &   & 2,0   \\ 
HD40335   &  SAO113311   & 05:58:13.52 $+$01:51:23.0 & $+$6&  $-$7 &   A0   &  6.441 
& 0.025 &  6.41 & 0.02 & 2,5 \\
BS2228    &  HD43244     & 06:17:34.65 $+$46:25:26.2 & $-$44 & $+$11 &   F0V   &       &      
&  5.87 & 0.03 & 0,2 \\ 
HD44612   &  SAO41080    & 06:24:46.60 $+$43:32:54.5 & $+$0&  $-$22 &   A0   &  7.050 
& 0.014 &  7.07 & 0.03 & 3,3\\
FS123     &  P486-R      & 08:51:11.88 $+$11:45:21.5 & $-$8&  $-$6 &   B8   & 
10.25\phantom {0} & 0.02 &       &   & 4,0   \\ 
HD77281   &  SAO136505   & 09:01:38.01 $-$01:28:34.8 & $-$16 & $-$13 &   A2   &  7.041 
& 0.014 &  7.02 & 0.03 & 5,5\\
FS125     &  P259-C      & 09:03:20.60 $+$34:21:03.9 & & &   G8   & 
10.33\phantom {0} & 0.03 &       &   & 3,0   \\ 
GL347A    &  G161-33     & 09:28:53.50 $-$07:22:15.0 & $-$165 & $-$672 &   M2.5V &  7.367 
& 0.014 &       &   & 3,0   \\
HD84800   &  SAO43050    & 09:48:44.64 $+$43:39:55.6 &  $-$28 &  $-$30 &  A2II   &  
7.547 & 0.013 &  7.56 & 0.02 & 5,5\\
FS129     &  LHS2397a    & 11:21:48.95 $-$13:13:07.9 & $-$399 & $-$348 &  M8V   & 
10.03\phantom {0} & 0.03 &       &   & 1,0   \\
HD105601  &  SAO62866    & 12:09:27.80 $+$38:37:54.6 &  $-$33 &  $-$60 &  Am    &  6.669 
& 0.014 &  6.70 & 0.02 & 3,3\\ 
HD106965  &  SAO119313   & 12:17:57.54 $+$01:34:31.1 & $-$26 &  $-$7 &   A2   &  7.311 
& 0.010 &  7.32 & 0.04 & 6,3\\
FS134     &  LHS2924     & 14:28:43.37 $+$33:10:41.5 & $-$337 & $-$747 &  M9V   & 
10.10\phantom {0}  & 0.02 &       &    & 3,0  \\ 
HD129653  &  SAO64289    & 14:42:39.56 $+$36:45:24.3 & $+$29 &  $-$10 &   A2   &  6.920 
& 0.014 &  6.99 & 0.02 & 3,3\\
HD129655  &  SAO140097   & 14:43:46.44 $-$02:30:20.0 &  $-$31 &  $-$26 & A2   &  6.666 
& 0.014 &  6.69 & 0.02 & 3,3\\  
HD136754  &  SAO83785    & 15:21:34.53 $+$24:20:36.1 &  $-$36 &  $-$9 & A0   &  7.158 
& 0.010 &  7.13 & 0.04 & 5,4\\
BS6092    &  HD147394    & 16:19:44.44 $+$46:18:48.1 & $-$13 &  $+$39 &  B5IV   &       &      
&  4.37 & 0.03 & 0,2\\
FS138     &  P275-A      & 16:28:06.72 $+$34:58:48.3 &  $-$12 &  $+$5 & A1   & 
10.44\phantom {0} & 0.03 &       &  & 3,0    \\ 
FS140     &  S587-T      & 17:13:22.65 $-$18:53:33.8 &  &  & G9   & 
10.34\phantom {0} & 0.03 &       &   & 2,0   \\
HD162208  &  SAO66344    & 17:47:58.56 $+$39:58:50.9 &  $-$12 &  $+$130 &  A0   &  7.125 
& 0.014 &  7.05 & 0.02 & 3,3\\ 
HD161903  &  SAO141886   & 17:48:19.22 $-$01:48:29.7 &  $+$8 &  $+$0 & A2   &  7.034 
& 0.014 &  6.97 & 0.02 & 3,3\\ 
HD161743  &  SAO209292   & 17:48:57.93 $-$38:07:07.5 & $+$3 &  $-$10 &   B9IV   &  
7.623 & 0.020 &  7.67 & 0.03 & 2,1\\
FS147     &  P230-A      & 19:01:55.27 $+$42:29:19.6 &  $-$1 &  $+$0 &  A0   &  
9.84\phantom {0} & 0.02 &       &  & 3,0    \\
GL748     &  G22-18      & 19:12:14.60 $+$02:53:11.1 & $+$1789 &  $-$520 &  M3.5V &  6.012 
& 0.020 &  6.00 & 0.03 & 2,2\\
FS148     &  S810-A      & 19:41:23.52 $-$03:50:56.1 & $-$1 & $-$4 &   A0   &  
9.46\phantom {0} & 0.02 &       &  & 4,0    \\  
FS149     &  P338-C      & 20:00:39.25 $+$29:58:40.0 & $+$5 &  $-$4 &   B7.5 & 
10.06\phantom {0} & 0.02 &       &   & 5,0   \\
BS7773    &  HD193432    & 20:20:39.82 $-$12:45:32.7 &  $+$16 &  $-$15 &  B9IV   &       
&      &  4.86 & 0.02 & 0,3\\
FS150     &  CMC513807     & 20:36:08.44 $+$49:38:23.5 & $+$8 & $+$9 &  G0   &  
9.91\phantom {0} & 0.02 &       &  & 4,0    \\
GL811.1   &  Wolf 896     & 20:56:46.60 $-$10:26:54.6 & $-$24 & $-$1110 &   M2.5V &  
6.691 & 0.014 &  6.72 & 0.02 & 3,3\\ 
HD201941  &  SAO126618   & 21:12:45.32 $+$02:38:33.9 &  $-$28 & $-$20 &  A2  &       &      
&  6.63 & 0.03 & 0,3\\
HD203856  &  SAO71278    & 21:23:35.53 $+$40:01:07.0 &  $+$29 & $+$7 & A0   &  6.871 
& 0.013 &  6.84 & 0.02 & 5,3\\
SAO34401  &  HD212533    & 22:23:42.24 $+$55:12:25.1 &  $+$13 & $-$1 &  F0V   &  7.735 
& 0.013 &  7.70 & 0.02 & 6,3\\
BS8541    &  HD212593    & 22:24:30.99 $+$49:28:35.0 &  $-$5 & $-$3 & B9Iab   &       &      
&  4.20 & 0.02 & 0,3\\
FS155     &  CMC516589      & 23:49:47.82 $+$34:13:05.1 &  &  & K5   &  
9.32\phantom {0} & 0.02 &       &   & 5,0   \\
\end{tabular}
\end{table}

\newpage

\begin{figure}
\psfig{file=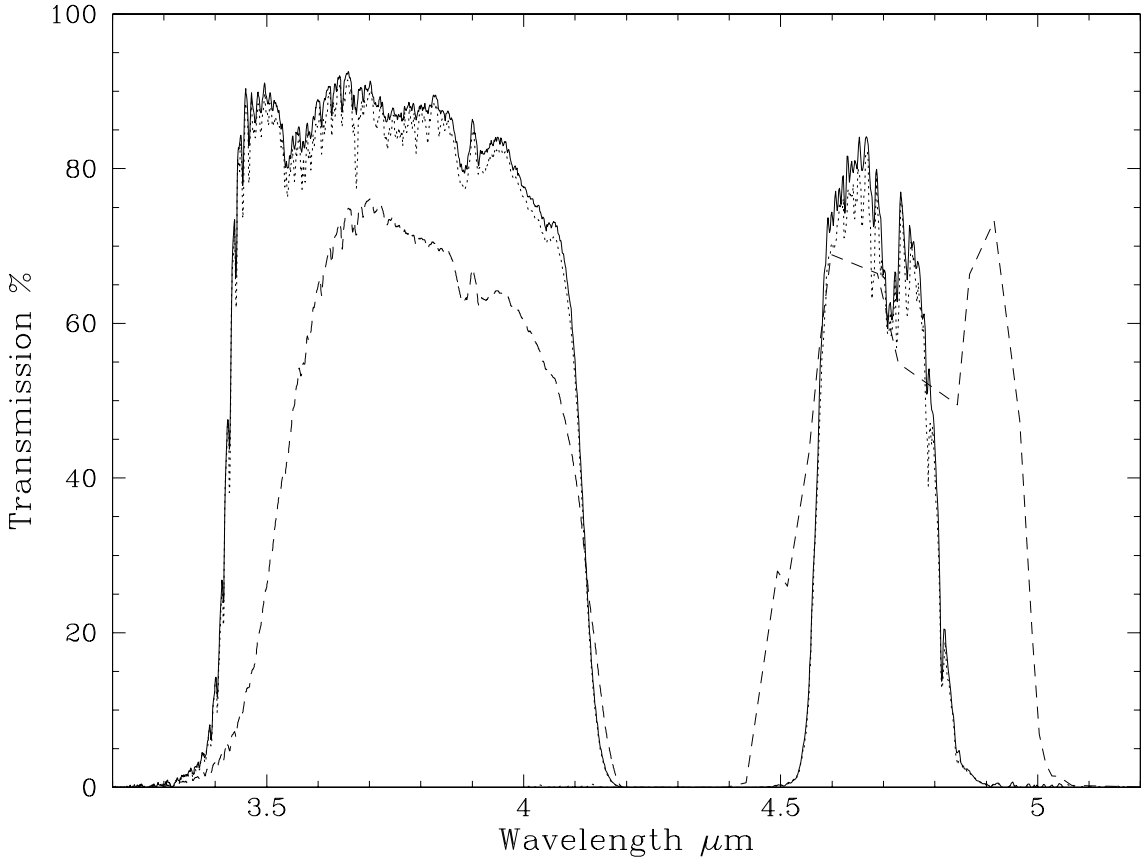,width=0.99\hsize,angle=-90}
\caption{\label{filters} Transmission profiles for the current
Mauna Kea consortium L$^{\prime}$ and M$^{\prime}$ filters
for cold, instrument, temperatures, including the effect
of absorption by the atmosphere.  Solid line is for dry conditions
with a precipitable water column of 1.2~mm, and the dotted line
for wetter conditions of 3~mm water vapour.  Other transmission
factors such as detector QE and telescope and instrument optics are 
effectively flat across
the bandpasses and are not included (see discussion in text).
The dashed line shows the profile for the previous
UKIRT system L$^{\prime}$ and broadband M filter.  
}
\end{figure}

\begin{figure}
\psfig{file=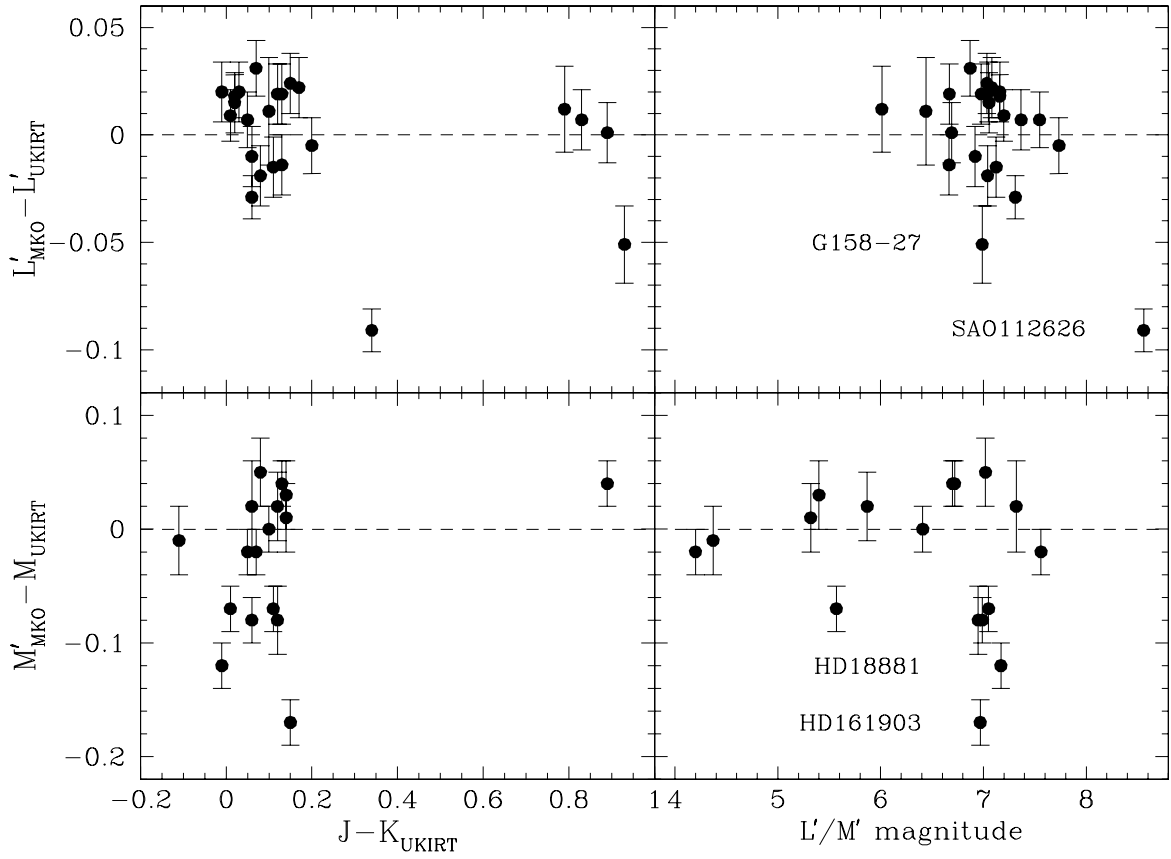,width=0.99\hsize,angle=-90}
\caption{\label{comps} Comparison of present results with previous
UKIRT system magnitudes, left panel as function of colour,
right panel as a function of brightness.  See text for discussion.
}
\end{figure}

\end{document}